\documentclass[11pt]{article}

\usepackage{amsmath}
\usepackage{graphicx}

\usepackage[figuresleft]{rotating}

\usepackage{hyperref}
\usepackage{algorithm2e}

\usepackage[defaultlines=3,all]{nowidow}

\usepackage{csquotes}

\begin{document}

\title{Real-Time Guarantees in Routerless Networks-on-Chip}

\author{Leandro Soares Indrusiak and Alan Burns}

\date{University of York}










\maketitle

\begin{abstract}
This paper considers the use of routerless networks-on-chip as an alternative on-chip interconnect for multiprocessor systems requiring hard real-time guarantees for inter-processor communication. It presents a novel analytical framework that can provide latency upper bounds to real-time packet flows sent over routerless networks-on-chip, and it uses that framework to evaluate the ability of such networks to provide real-time guarantees. Extensive comparative analysis is provided, considering different architectures for routerless networks and a state-of-the-art wormhole network based on priority-preemptive routers as a baseline.
\end{abstract}

\section{Introduction}\label{intro}

On-chip interconnects have played a key role in the performance and scalability of multicore and many-core processor architectures over the past two decades, and routerless networks-on-chip NoCs are one of the latest developments in that area. The concept of routerless NoCs builds on two well-studied interconnect features: ring topology and deflection routing. \emph{Ring topologies} for on-chip interconnects have been studied in academic research~\cite{Tortosa2002}~\cite{Bourduas2007} and are widely used in commercial processor interconnects like the Intel Sandy Bridge~\cite{IntelSandyBridge2012} and the IBM Cell~\cite{IBMCell2006}. In those cases, data is sent over one or more multi-hop rings towards its destination (typically over the shortest path). \emph{Deflection routing} exploits the possibility of circular routes over ring topologies, allowing data to take longer paths and circle around the ring interconnect one or more times instead of sitting in buffers in the case of link contention~\cite{Gomez2008}~\cite{Ausavarungnirun2014}. The application of those two features to NoCs allowed for lower hardware overhead, because of the limited need for buffering and flow control circuitry~\cite{Moscibroda2009}. The advantages with regards to energy dissipation are less clear-cut, but experimental work has shown that typically there are moderate amounts of energy savings despite longer routes due to deflection ~\cite{Michelogiannakis2010}. 

Recent works on the so-called \emph{routerless NoC} architectures~\cite{alazemi_routerless_2018}~\cite{liu_imr_2016}~\cite{xiao_onion_2019}~\cite{lin_deep_2020} have pushed those concepts one step further and advocated for additional savings in hardware overheads by completely removing the routing function from the NoC interconnect. In previous ring-based defletion-enabled NoCs, data packets would be injected by the source processing core via a network interface, and the network would route them across one or more rings, deflecting them as needed when contention arises, until they reach their destination core. Those architectures would still require some sort of buffering and flow control mechanisms to enable the transfer of packets from one ring to another, as well as routing mechanisms on every network hop along the way. Routerless NoCs do not allow packets to change rings along their way, removing completely the need for routing and requiring only minimal buffering and flow control on each network switch. To enable full connectivity among processing cores, routerless NoCs have carefully selected topologies with multiple rings that, among them, connect all cores of the system. Packets are therefore always injected into a ring that passes through their source and their destination cores. The only routing decision that still must be performed is the choice of ring that should be used given the destination of each data packet, and there may be more than one possibility, depending on the topology. That decision is made at the interface of the network before the packet is injected into a ring, typically through a routing table. Figure \ref{fig:overview} provides an overview of a routerless NoC, and more details about the architecture will be given in subsection \ref{routerless_architecture}.

The advantages of routerless NoCs in terms of performance, hardware area and energy overheads have been experimentally evaluated in ~\cite{liu_imr_2016} and ~\cite{alazemi_routerless_2018}. With regards to performance, Liu et al. ~\cite{liu_imr_2016} reported improvements of more than 200\% in saturation throughput and more than 70\% in average latency when comparing their IMR (Isolated Multi-Ring) routerless NoC against several router-based NoC architectures including mesh and torus networks. The area and energy savings they reported were modest, never reaching more than 10\% improvement over the baselines.  Following a more aggressive reduction of buffering circuitry, Alazemi et al.~\cite{alazemi_routerless_2018} reported more than 10x reduction in power consumption compared to a router-based mesh NoC (and nearly 8x improvement against Liu et al.'s IMR),
as well as area savings of approximately 85\% and 70\% compared with a router-based mesh NoC and IMR, respectively.

Given the figures reported by ~\cite{liu_imr_2016} and ~\cite{alazemi_routerless_2018}, it is likely that the resource efficiency of routerless NoC architectures would be beneficial to embedded and real-time multiprocessor platforms. However, none of the works on routerless NoCs addresses the problem of packet latency bounds, focusing instead on average-case latency metrics. In this paper, we present the first approach to obtain latency upper bounds to real-time packet flows sent over routerless networks-on-chip. With that, we aim to enable the use of routerless NoCs as the communication backbone for real-time and embedded multiprocessors. Furthermore, we aim to shed light on the inherent trade-offs posed by such networks and show in which circumstances should designers replace traditional router-based networks by routerless NoCs, and in which they should not.  

The paper is organised as follows: Section \ref{background} provides an overview of the state-of-the-art in real-time networks-on-chip, and provides a more detailed background on routerless NoCs; Section \ref{analysis} presents a novel analytical framework that can provide latency upper bounds to real-time packet flows sent over routerless NoCs, which is the first contribution of this paper; the second contribution of the paper, in Section \ref{experiments}, addresses the use of the proposed analytical framework to evaluate the ability of routerless NoCs to provide real-time guarantees under different configurations, compared against a state-of-the-art router-based NoC; the paper is then closed with a summary of the insights uncovered by the experimental work and with numerous lines of further research that were opened by the proposed framework. 

\section{Background}\label{background}

\subsection{Real-time Wormhole Networks-on-Chip}\label{wormhole}

Wormhole switching has been widely used in NoCs because of its balance between performance and buffering overheads in the router. Unlike switching protocols such as store-and-forward (SAF) and virtual cut-through (VCT), wormhole does not require buffers to have capacity to store a full packet. Each packet is forwarded as a sequence of fixed size data units (flits), the first of them (the header) carries information about routing and packet size. As the header advances along the specified route, the remaining flits follow in a pipelined way. If the output link requested by a header flit is supposed to be routed via a link already in use, it is blocked until the link becomes available. In this situation, the second flit will then be blocked by the first one, and so on, until all flits stall. They will then remain buffered in one or more routers along the packet route (depending on the buffering capabilities of each router) until the header is released, so the pipelined transmission can continue.

The smaller the buffers on each router, the larger the number of routers that will have to store a given packet in a blockage scenario. If there is not enough buffer space distributed over routers in the packet route, the backpressure will propagate back to the packet sender, preventing it from injecting further flits into the network. Since a packet can be stored by several routers and occupy multiple links at a time, the potential congestion over the network is increased. This makes it harder to predict the time it takes for a given packet to cross the network, because many of the links along its route may be blocked by other packets. This is not the case in store-and-forward (SAF) switching, where each packet uses only one link at a time, or in virtual cut-through (VCT), where packets are only stored in the router where they experience blocking~\cite{Dally04}. Wormhole switching has been widely preferred over SAF or VCT in on-chip interconnects because having smaller buffers allowed for smaller overheads in silicon area and energy dissipation. 

Several arbitration and flow control mechanisms were proposed to improve time predictability in on-chip networks, using resource sharing policies such as time-division multiplexing (TDM)~\cite{Goossens05} and prioritised virtual channels (VCs)~\cite{Bolotin04}. TDM tries to avoid latency interference between packets by reserving link bandwidth to each packet flow. Priority-arbitrated virtual channels allow packets to interfere with each other but aims to quantify the interference upper bounds for each packet flow. Priority-based mechanisms are seen as superior, since they are work conserving (do not reserve resources) and more flexible (do not require exact knowledge of packet sizes or injection times), but not always preferred because of higher overheads in area and energy dissipation. A wider review of real-time network-on-chip architectures is beyond the scope of this paper, and a good survey can be found in ~\cite{Hesham2017}. Our focus in this paper is actually on the analytical models that evaluate how well a given architecture can provide guarantees to real-time application communication traffic. That aspect is not covered in detail by ~\cite{Hesham2017}, so we provide a brief review and additional references over the next paragraph, with a focus on models for priority-based NoCs.

Most of the real-time analytical models for priority-based wormhole NoCs have been based on analysis developed in the mid 1990s for general purpose wormhole networks ~\cite{Mutka94}~\cite{Hary97}~\cite{Kim98}. NoC analysis models proposed by Shi and Burns~\cite{Shi08} and Kashif and Patel~\cite{Kashif2016} customised those general models for the specifics of the on-chip communication mechanisms and were widely used until the discovery of the multi-point progressive blocking (MPB) problem by Xiong \textit{et al.}~\cite{Xiong16}. MPB exposed a flaw in the assumptions of the original analysis approaches from the 1990s which made all subsequent analytical models unsafe. The problem manifests itself in indirect interference scenarios happening downstream from the path of a given packet, but affecting that packet in complex ways due to backpressure effects caused by the limited buffering per router (which is a key part of wormhole switching). As MPB was underestimated in all those analytical models, they can lead to optimistic predictions of schedulability. Since the discovery of the MPB problem, it has been safely modelled by several approaches ~\cite{Xiong17}~\cite{Indrusiak2018}~\cite{Nikolic2018} (i.e. no known optimistic counter-examples). Despite the consistent increase in tightness from one approach to the next, all of them are still significantly more conservative than previous analyses that did not model MPB. In ~\cite{Giroudot2019}, authors used network calculus in an attempt to improve the tightness against ~\cite{Nikolic2018}, which is based on response-time analysis as its predecessors, but they could not show any dominance of one model over the other. This provided more evidence that the MPB problem is not an artifact of a specific type of analysis, but instead is an inherent issue that comes with wormhole networks and backpressure (albeit a rare one - it took more than a decade for it to be uncovered). Accordingly, some of the latest works in the area of real-time NoCs are attempts to avoid or control backpressure as a way to prevent MPB, but they either rely on global wires which may prevent an efficient and reliable implementation in silicon ~\cite{NikolicECRTS2019}~\cite{Ueter19}~\cite{Gonzalez2020} or require complex memory management in every NoC core ~\cite{Burns2020}.

\subsection{Routerless Networks-on-Chip}\label{routerless_architecture}
Recent works have proposed network architectures that completely remove flow control, routing and/or switching logic, aiming to reduce the hardware and energy overheads in network routers. Processing cores are still laid out in 2D grids and network links only connect neighbouring cores, avoiding global wire issues and allowing for regular structures that can be reliably implemented in silicon. 

In ~\cite{Ausavarungnirun2014}, authors use a hierarchical ring topology with no in-ring buffering or buffered flow control. Packets are allowed to transit across rings, so inter-ring buffers are employed. These are limited, forcing packets to keep circling the ring until the buffer has enough space. 

Liu et al.~\cite{liu_imr_2016} proposed IMR, which uses an even more economic design where packets are not allowed to transit across rings, so the problem shifts to the efficient definition of rings that can provide full connectivity and low average hop  count between communicating cores. The authors propose a genetic algorithm formulation based on a clever bit-stream representation of potential rings, and use it to evolve ring configurations that minimise the overall number of interconnect links, average latency and hop count, while ensuring full network connectivity. IMR uses full-packet buffers on each switch for each ring, providing temporary storage in case a packet is not granted arbitration to the output link, which avoids sophisticated flow control and completely prevents backpressure. Each switch has a single ejection link connecting it to the local processing core, which is shared by all rings passing through that switch. Due to the lack of flow control and limited buffering capabilities, a packet may be forced to circle around the loop repeatedly if the ejection link to their destination core is not available. To prevent livelock (i.e. packets looping around the ring forever), all packets are timestamped so that the ejection links can be arbitrated using an Oldest-First policy. 

Alazemi et al. ~\cite{alazemi_routerless_2018} proposed a number of improvements to IMR: a constructive heuristic called RLrec to efficiently define and place rings ensuring full communication across the network; a protocol to share full-packet buffers among multiple rings going through the same switch; a possibility of multiple ejection links; and a different livelock prevention mechanism based on a circling counter. The authors applied for a patent for this design, which was granted in 2020 ~\cite{routerlesspatent}.

\begin{figure}[h]
  \centering
  \includegraphics*[scale=1.0]{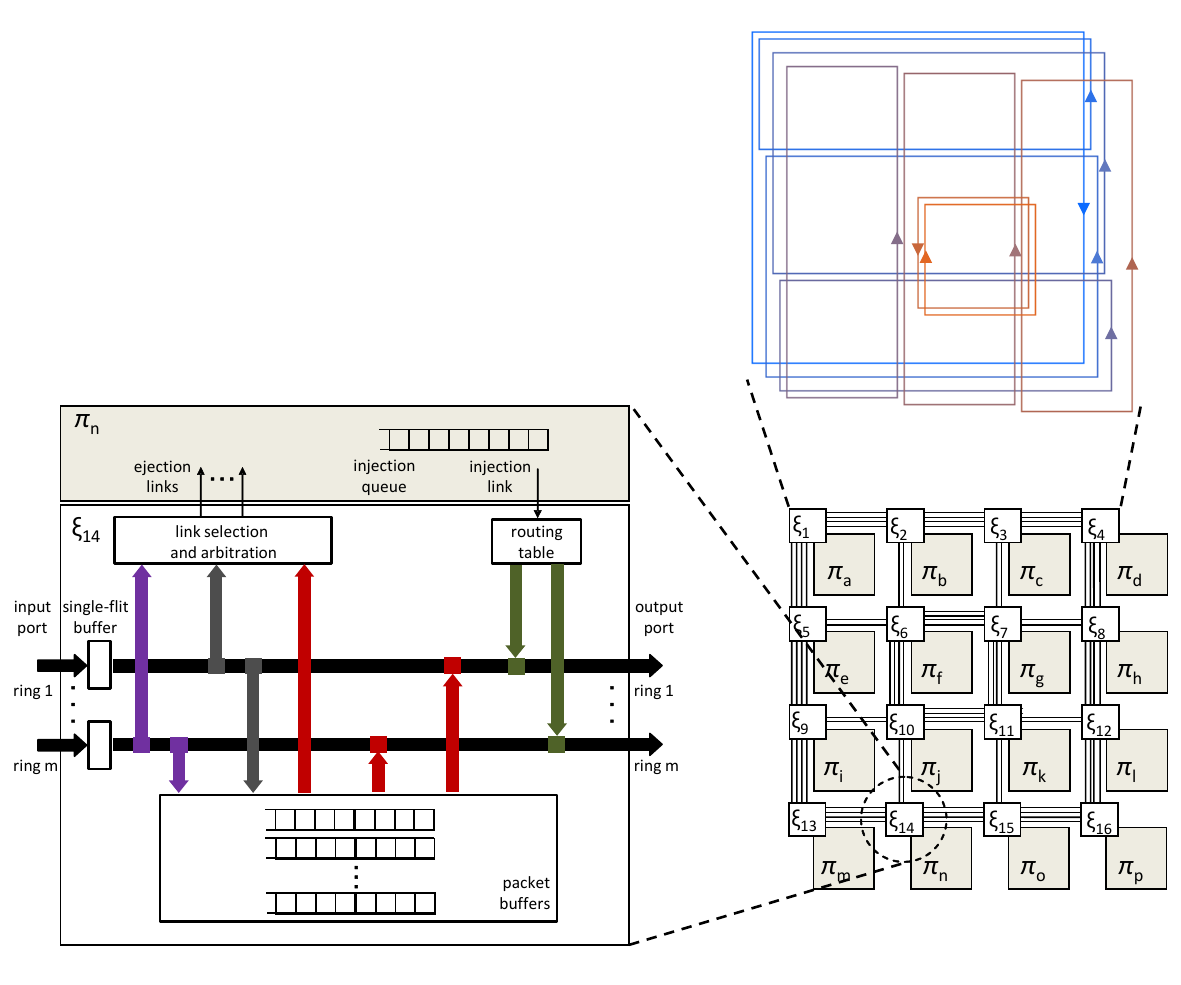}
  \caption{Routerless network switch (adapted from ~\cite{alazemi_routerless_2018}), 4x4 network, and detail of the topology generated by the RLrec heuristic.}
  \label{fig:overview}
\end{figure}

Figure \ref{fig:overview}, adapted from ~\cite{alazemi_routerless_2018}, shows the main components of a routerless network as proposed in ~\cite{liu_imr_2016} and ~\cite{alazemi_routerless_2018}. The lower-right part of the figure shows a 16-processor network organised in a 4x4 grid, each processor connected to a switch. The switches are interconnected by 10 rings following the RLrec approach from ~\cite{alazemi_routerless_2018} (detail in the upper-right part of the figure). One of the switches is shown in more detail in the left part of the figure, including the injection and ejection links used by the local core to connect to the network, the input and output ports for each ring passing through the switch, and the full-packet buffers used for temporary storage. Packets arriving from the ring are transferred into the switch via the ring input port, at a rate of one flit per network cycle (which correspond to one clock cycle in the hardware implementations in ~\cite{liu_imr_2016} and ~\cite{alazemi_routerless_2018}). In the remainder of this paper, we will use a network cycle as the unit for all time-related parameters and metrics, and will refer to it simply as `cycle'. During a network cycle, the following operations can happen:

\begin{itemize}

\item Flits arriving at input ports of each ring are stored in their respective flit buffers.
\item Flits that had arrived in the previous cycle are transferred from the flit buffers to an ejection link (if their destination is the local processor), to the respective ring's output port (if there is no ongoing packet payload injection or buffer draining in their ring) or to a packet buffer (if there is an ongoing packet payload injection or buffer draining in their ring).
\item Header flit of the packet at the head of the injection queue is transferred to the output port of the ring indicated by the routing table, but only if the packet buffer and flit buffer of that ring are empty (i.e. ring traffic always has precedence over packet header injection).  
\item Payload flit of the packet at the head of the injection queue is transferred to the output port of the same ring that the header or payload flit of that same packet has used during the previous cycle (i.e. payload flit injection always has precedence over ring traffic to avoid packet interleaving).
\item Flits stored in the head of packet buffers are transferred to the output port of their respective rings if there is no ongoing packet payload injection. 
\end{itemize}

As one can infer from the operations listed above, a routerless network requires no flow control. A switch can always send a flit per cycle to the downstream switch of each of its rings, as it can be sure that there is space in the downstream flit buffer since any flit sent over the previous cycle would by then be transferred further to an output port, to a packet buffer, or to an ejection link.

From the point of view of the output port, the protocol is straightforward: it outputs a flit, if one exists, from the packet buffer, flit buffer or injection queue, in that order of priority unless there are additional flits to be injected (as packet injections are non-preemptive).

\section{Real-Time Analysis for Routerless Networks-on-Chip}\label{analysis}
This section presents the first real-time analysis model for a routerless NoC, initially by assuming exclusive injection and ejection links for each ring, and then by analysing the impact of sharing injection and ejection links among multiple rings.

\subsection{System Model}\label{system_model}
Let us model a routerless network-on-chip such as the ones discussed in subsection \ref{routerless_architecture} as a set of processing cores $\Pi = \{\pi_a,\pi_b, \ldots, \pi_z\}$; and a set of rings $O = \{o_1,o_2, \ldots, o_m\}$
where, representative ring $o$ is defined by the ordered set of $r^o$ switches $\Xi^o = \{\xi_1,\xi_2, \ldots, \xi_{r^o}\}$
and the
set $\Lambda^o$ of $3r^o$ unidirectional links 
which includes a link between subsequent switches of the ring and two links to and from the processing core connected to each switch (referred to as the ejection and injection links). Each switch $\xi$ of $\Xi^o$ includes a buffer of size $B^o$ which must be able to store even the largest packet sent to its respective ring.

It is possible (and likely, in the topologies proposed in ~\cite{liu_imr_2016} and ~\cite{alazemi_routerless_2018}) that a switch is part of more than one ring, e.g. $\xi_2 \in \Xi^{o^1}, \xi_2 \in \Xi^{o^3}, \xi_2 \in \Xi^{o^7}$. In such a case, the switch encompasses the links and buffers of all rings going through it (as shown in Figure \ref{fig:overview}). 

To model the traffic load injected to the network, we define a set $\Gamma$ of $n$
real-time traffic-flows  (or just \emph{flows} for short) $\Gamma$ =$ \{\tau_1,
\tau_2, \ldots \tau_n\}$.
Each flow $\tau_i$ gives rise to a potentially unbounded sequence
of \emph{packets} that are sent over the network \emph{flit-by-flit}. The flow has a set of properties and timing
requirements which are characterised by a set of attributes:
$\tau_i$ = ($T_i$, $D_i$, $L_i$, $J_i$, $\pi_i^s$,
$\pi_i^d$). The source and destination cores of a flow are denoted respectively by $\pi_i^s$ and $\pi_i^d$.
We assume that all the flows which require timely delivery are either
periodic or sporadic, and the lower bound interval on the time between
releases of successive packets is called the period ($T_i$) for the
flow, measured in cycles. 

The maximum packet size of a flow, in flits, is denoted by $L_i$. As a ring forwards one flit per cycle, $L_i$ is also the time it takes for a packet of $\tau_i$ to completely cross a link, in cycles. Each real-time flow has a relative deadline ($D_i$) which is the upper bound restriction on network latency. In this work we assume for simplicity that
$D_i \le T_i$. Finally, we assume that the maximum deviation of successive packet releases from the flow's period, i.e. the maximum packet jitter, is known and represented as $J_i$. $D_i$ and $J_i$ are also given in cycles.

We now define as $\Gamma^o$ a subset of $\Gamma$ with all the flows using ring $o \in O$. Each traffic flow $\tau_i$ is bound to a single ring $o \in O$, so all possible $\Gamma^o$ subsets of $\Gamma$ are mutually exclusive. Similarly, we define $\Gamma^o_\xi$ as the subset of $\Gamma^o$ with all the flows that use switch $\xi$ of ring $o$. As it is possible, and perhaps likely, that a flow $\tau$ goes through several of the switches of its ring, the subsets $\Gamma^o_\xi$ of $\Gamma^o$ are not necessarily mutually exclusive.
To be more precise about the nature of the flows in $\Gamma^o_\xi$, we divide them into three mutually-exclusive subsets:
$\Gamma^o_{\xi in}$ (packets flowing into $\xi$ via its injection port and exiting via the output port),
$\Gamma^o_{\xi out}$ (packets flowing into $\xi$ via its input port and exiting via the ejection link) and
$\Gamma^o_{\xi thru}$ (packets flowing into $\xi$ via its input port and exiting via the output port).
In the following, we will further refine such subsets to precisely represent resource sharing between flows,
and from that we will qualify the potential timing interference between them.

A main goal of this paper is to quantify the worst-case latency (response-time), $R_i$, of any packet of a given packet flow $\tau_i$. In the following subsections, we propose a framework of analytical models accounting for the contributions of the packet's own latency as well as upper-bounds on the latency interference from other packets sharing the same network resources (i.e. contention), for different configurations of the network. Once we are able to calculate its worst-case latency $R_i$, we can determine if a packet flow $\tau_i$ is \emph{schedulable}: if $R_i \le D_i$ then all packets of $\tau_i$ will always reach their destination before their respective deadline, even when encountering worst-case network contention. Generalising the same concept, we deem that a flowset $\Gamma$ to be schedulable when all flows $\tau_i \in \Gamma$ are schedulable.  

To support the proposed analytical framework, we also need to define specific subsets of the ordered set of switches $\Xi^o$ of a given ring. We therefore define the function $path^o(\pi_\alpha, \pi_\omega)$ to denote the ordered subset of $\Xi^o$ with the switches in the path between cores $\pi_\alpha$ and $\pi_\omega$; and the function $dpath^o(\pi_\alpha, \pi_\omega)$, which denotes the downstream path between those processing cores i.e. the exact same ordered subset of $\Xi^o$ except for the first switch (i.e. the one directly connected to $\pi_\alpha$). We then define that the absolute value of such a function denotes the number of switches in the respective path, so it should be clear that $| path^o(\pi_\alpha, \pi_\omega) | = | dpath^o(\pi_\alpha, \pi_\omega) | + 1$. 

We also define two secondary parameters for each traffic flow $\tau_i$: $C_i$ and $C^o_i$. $C_i$ represents the maximum transmission latency of a packet of $\tau_i$ from $\pi_i^s$ to $\pi_i^d$ when no contention exists. It is a parameter found in most real-time analysis for networks and is often called maximum basic latency or maximum no-load latency of a packet flow. Assuming that, in the absence of contention, a packet flit is transmitted per cycle by the ring, the value of $C_i$ in cycles can be obtained by adding the number of ring links between $\pi_i^s$ and $\pi_i^d$ (including injection and ejection ones) and the number of payload flits of the packet, which is given by $L_i-1$. The first term of that sum (given below) represents the time taken by the header flit to travel from the source to the destination, and the second term is the time it takes for all payload flits to arrive after the header, in pipeline fashion. The other secondary parameter $C^o_i$ represents the maximum deflection latency of a packet of $\tau_i$ when traveling
all the way around around ring $o$ when no contention exists. We therefore refer to it as maximum basic loop latency or maximum no-load loop latency. Following a similar intuition, $C^o_i$ can be found by the sum of the number of links in ring $o$ (given by $r^o +1$, where $r^o$ is the number of switches
in the ring) and the number of payload flits of $\tau_i$ (given by $L_i-1$).

\begin{equation}\label{eq:ci}
C_i \ =  |path(\pi_i^s,\pi_i^d)| + L_i - 1
\end{equation}

\begin{equation}\label{eq:cio}
C^o_i \ =  r^o + L_i
\end{equation}

\subsection{Basic Analysis}\label{basic_analysis}

To calculate the worst-case latency $R_i$ experienced by any packet of a packet flow $\tau_i$, we must analyse all sources of timing interference that can affect it. In this paper, we start with a set of assumptions that reduces the potential sources of timing interference. Once we are able to derive analytical models that can calculate latency upper-bounds under the initial set of assumptions, we will then progressively lift the assumptions and derive more general and widely applicable models.

Our first step assumes that rings are completely independent, meaning that their switches, buffers and links are not shared with other rings. In that case, packets of a flow $\tau_i$ injected to a ring $o$ can only suffer interference from the following sources:

\begin{enumerate}

\item Packets from any other previously injected flow $\tau_j \in \Gamma_{up_i}^{o}$ (where $\Gamma_{up_i}^{o}$ is the set $\Gamma^{o}_{\xi thru}$ with all flows going through the switch where the packets of $\tau_i$ are injected), as the packets of $\tau_i$ cannot start injection unless there are no flits flowing through that switch's output port. We refer to these sources of interference as \textbf{upstream direct interference}, or $\tau_j^{up}$ for short, because the packets of those flows must be injected in switches upstream from the switch where $\tau_i$ is injected, they must be injected before the injection of $\tau_i$'s packets, and will compete directly with $\tau_i$'s packets for the output link of the switch where $\tau_i$ is injected.

\item Packets from any other flow $\tau_j \in \Gamma_{down_i}^{o}$ (where $\Gamma_{down_i}^{o}$ is the subset of $\Gamma^{o}$ with all flows injected or buffered in any of the downstream switches along $\tau_i$'s path towards its destination), as they may (injected) or will (buffered) take precedence over $\tau_i$ in the access of the switch's output link, forcing it to be buffered in that switch until the transfer of the interfering packet is completed.  We refer to these sources of interference as \textbf{downstream direct interference}, or $\tau_j^{down}$ for short, because the packets of those flows will compete directly with $\tau_i$'s packets for the output link of switches that are downstream from the switch where $\tau_j^{down}$ is injected.  

\item Packets from any other flow $\tau_j \in \Gamma_{in_i}^{o}$ (where $\Gamma_{in_i}^{o}$ is the set $\Gamma^{o}_{\xi in}$ with all flows injected into the same switch as $\tau_i$). We assumed at this stage that each ring has its exclusive injection link, but such an assumption does not prevent it from being used by many local flows, so we must account for the interference they can cause. We refer to these sources of interference as \textbf{injection direct interference}, or $\tau_j^{in}$ for short, because the packets of those flows will compete directly with $\tau_i$'s packets for the injection link.

\item Packets from any other flow $\tau_k \in \Gamma_{upind_i}^{o}$ (where $\Gamma_{upind_i}^{o}$ is the subset of $\Gamma^{o}$ with all flows with a path that does not share links with the path of the packets of $\tau_i$, but causes direct interference to a flow that in turn causes upstream direct interference to $\tau_i$). This type of interference has been widely studied in wormhole networks, and is referred to as indirect interference in ~\cite{Shi08} and \textbf{upstream indirect interference} in ~\cite{Nikolic2018}. In effect, upstream indirect interference can amplify the perceived jitter of flows causing upstream direct interference, so this must be accounted for in the analysis.  

\end{enumerate}

Notice that downstream indirect interference (which is the cause of the MPB problem reviewed in subsection \ref{wormhole}) does not appear as a fifth item in the list discussed above. This is because the flow control and arbitration mechanisms in routerless NoCs do not allow for backpressure: if a packet does not gain arbitration to an output link of a switch, its flits will be completely buffered in that one switch, and will not prevent the progress of the flits that may be in-route over other upstream switches. This behaviour follows the principles of a virtual cut-through network, rather than the wormhole switching typically used in NoCs, and prevents all interference scenarios resulting from downstream indirect interference, including complex and rare ones such as MPB.

\begin{figure}[h]
  \centering
  \includegraphics*[scale=1.1]{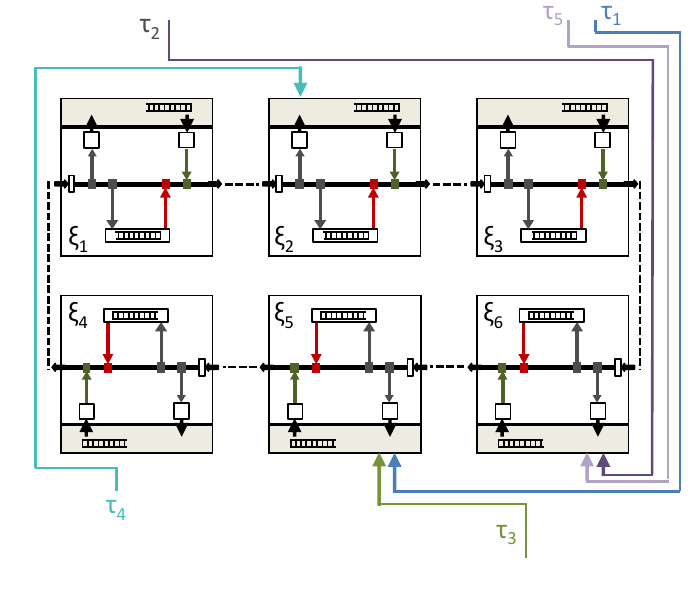}
  \vspace{-10pt}
  \caption{Scenario with five flows over a six-switch (clockwise) ring.}
  \label{fig:scenario}
\end{figure}

Figure \ref{fig:scenario} illustrates all sources of interference listed above in a scenario where five traffic flows are transferred over a six-switch (clockwise) ring $o$:  $\Gamma^{o}=\{\tau_1, \tau_2, \tau_3, \tau_4, \tau_5\}$. Let us first concentrate on $\tau_1$, which is injected into switch $\xi_3$. Before it can be injected, $\tau_1$ can suffer upstream direct interference from $\tau_2$, i.e. $\Gamma_{up_1}^{o} = \{\tau_2\}$. Once it is injected, it can also suffer downstream direct interference from $\tau_3$, i.e. $\Gamma_{down_1}^{o} = \{\tau_3\}$. Since it is injected into the same switch as $\tau_5$, it can suffer injection direct interference from it, i.e. $\Gamma_{in_1}^{o} = \{\tau_5\}$. Finally, we can also see that $\tau_1$ can suffer upstream indirect interference from $\tau_4$, because $\tau_4$ does not share any links in $\tau_4$'s path, but it directly interferes with $\tau_2$, which in turn causes upstream direct interference to $\tau_i$, i.e. $\Gamma_{upind_1}^{o} = \{\tau_4\}$. Following the same principles, we summarise in Table \ref{tab:scenario} the different sources of interference for all five flows in this scenario.

\begin{table}[h]
\centering
\caption{Sources of interference for all five flows in Figure \ref{fig:scenario}.}
\label{tab:scenario}
\begin{tabular}{|l|c|c|c|c|}
\hline
\textbf{$\tau_i$} & \textbf{$\Gamma_{up_i}^{o}$} & \textbf{$\Gamma_{down_i}^{o}$} & \textbf{$\Gamma_{in_i}^{o}$} & \textbf{$\Gamma_{upind_i}^{o}$} \\ \hline
$\tau_1$            & \{$\tau_2$\}           & \{$\tau_3$\}             & \{$\tau_5$\}           & \{$\tau_4$\}              \\ \hline
$\tau_2$            & \{$\tau_4$\}           & \{$\tau_1$, $\tau_5$\}          & $\emptyset$       & $\emptyset$          \\ \hline
$\tau_3$            & \{$\tau_1$\}           & $\emptyset$         & $\emptyset$       & \{$\tau_2$, $\tau_5$\}           \\ \hline
$\tau_4$            & $\emptyset$       & \{$\tau_2$\}             & $\emptyset$       & $\emptyset$          \\ \hline
$\tau_5$            & \{$\tau_2$\}           & $\emptyset$         & \{$\tau_1$\}           & \{$\tau_4$\}              \\ \hline
\end{tabular}
\end{table}

Let us now consider the four sources of interference we identified above, and derive an analytical model that provides a safe bound to the worst-case latency $R_i$ of a flow $\tau_i$. By looking carefully at the nature of those sources of interference, we can claim that sources listed under items 1, 3 and 4 can only interfere with $\tau_i$ before its injection, while sources under item 2 can only interfere after its injection. This allows us to break the interference analysis into two completely separated stages: worst-case interference before injection $I_i^{pre}$ and worst-case interference after injection $I_i^{pos}$.

The \textbf{worst-case interference before injection} $I_i^{pre}$ is the largest amount of time a packet of $\tau_i$ must experience before it can have access to the output link of the switch it is injected into. We argue that this is equal to the longest busy period of that output link, as it serves newly injected packets to the switch (injection direct interference, item 3), packets arriving from the switch's input and previously buffered packets (upstream direct interference, item 1). As soon as the output link becomes idle, it is immediately acquired by $\tau_i$, which will then transmit the whole packet to the next downstream switch without experiencing any additional injection, upstream direct or indirect interference. Eq. \ref{eq:ipre} provides an upper bound for $I_i^{pre}$:

\begin{equation}\label{eq:ipre}
I_i^{pre} \ =  1 + \sum_{\tau_j \in \Gamma_{in_i}^{o} } {L_j} + \sum_{\tau_j \in \Gamma_{up_i}^{o} } { \left\lceil
{\frac{I_i^{pre} + J_j  + J_j^k }{T_j} } \right\rceil \cdot L_j }
\end{equation}

The third term of Eq. \ref{eq:ipre} follows a classic formulation of a busy period, with the term inside the ceiling function representing the maximum number of packets of $\tau_j^{up}$ transferred through the output port of the switch during an arbitrary time window with length equal to the numerator of that term. A recurrence relationship ensures that the considered time window is expanded by all terms of the equation. The third term also considers the maximum release jitter of the interfering packets $J_j$, as in ~\cite{Tindell1994}, and the maximum jitter amplification due to upstream indirect interference $J_j^{k}$, as in ~\cite{Shi08}. The choice of a safe value for $J_j^{k}$ in routerless NoCs is slightly more complicated than in the case addressed in ~\cite{Shi08}, so we will address this issue in more detail in subsection \ref{indirect}.

The second term of Eq. \ref{eq:ipre} accounts for the interference caused by $\tau_j^{in}$ flows. Given that in a schedulable system each flow can have at most one packet in-route (i.e. $D \le T$), only a single packet of each of the $\tau_j^{in}$ flows can interfere on a given release of $\tau_i$. For each of those packets, the interference happens when $\tau_j^{in}$ is in the injection queue when $\tau_i$ is released (so $\tau_i$ must wait in the queue), or when $\tau_j^{in}$ has been totally or partially injected before $\tau_i$'s release and its injection has caused the buffer to fill up (and therefore it must be drained before $\tau_i$ can be injected). In either case, the upper-bound to the interference is the maximum packet size $L$ for each of the flows. Notice that we do not double-account the interference of $\tau_j^{in}$ packets over the queue or the switch buffer: for a given $\tau_j^{in}$ flit, it will either be in the queue by the time $\tau_i$ is released or it will have already been injected and thus its interference will be in the form of buffer draining (i.e. upstream packets that it may have forced to be buffered). So the maximum interference caused by a $\tau_j^{in}$ packet will always be $L_j$ regardless if its flits are all in the queue by the time $\tau_i$ is released, or if they have all been injected and caused the buffer to fill up, or if part of them are in the queue and part of them have been injected (and partially filled up the buffer).

Due to the periodic nature of packet flows, it is possible that the busy period calculated by the second and third terms of Eq. \ref{eq:ipre} could repeat itself again and again, one immediately after another, completely preventing the injection of new packets into the ring. By adding the first term to the equation, we guarantee that there will be at least a single free cycle following the busy period calculated by the equation, thus allowing for the injection of a new packet.

Now, let us focus on the second stage of the analysis and formulate the \textbf{worst-case interference after injection} $I_i^{pos}$. Once a packet is injected into the ring, it can only suffer downstream direct interference from packets injected or buffered in the switches along its way towards its destination. More precisely, only a single packet of a single $\tau_j^{down}$ flow can cause interference per switch crossed by $\tau_i$. Either that packet is a newly injected packet, which will then force the buffering of $\tau_i$ until it completes the injection; or that packet is a previously buffered packet that is flowing out of the switch as $\tau_i$ arrives. Regardless of the scenario, $\tau_i$ will gain arbitration to the output link as soon as the interfering packet finishes its transmission. Therefore, an upper bound to the interference suffered by $\tau_i$ after its injection is given by Eq. \ref{eq:ipos}:

\begin{equation}\label{eq:ipos}
I_i^{pos} \ =  |dpath^o(\pi_i^s,\pi_i^d)|  \cdot B^o 
\end{equation}

As defined in subsection \ref{system_model}, the function \emph{$dpath^o$} provides the set of switches in ring $o$ the packets must go through between processing cores $\pi_i^s$ and $\pi_i^d$ (i.e. the source and destination cores of flow $\tau_i$) except for the injection switch (as the interference over that switch is completely accounted in $I_i^{pre}$). The absolute value of that function provides the number of switches in that set, which is then multiplied by $B^o$ (the size of the packet buffer in every switch of the ring, also defined in subsection \ref{system_model}), which is an upper-bound to the number of flits can be either injected or buffered in each of those switches before $\tau_i$ can acquire the output port. 

While clearly safe, we can argue that using $B^o$ as an upper-bound to the interference over each and every switch along the downstream path of a packet can be pessimistic. Given the maximum sizes of the packets injected into each switch along the downstream path, it is likely that many of their buffers may never be completely full. We therefore define $B^o_\xi$ as the maximum number of flits that will be stored in the packet buffer of the switch $\xi$ of ring $o$:

\begin{equation}\label{eq:maxbuffer}
B^o_\xi \ =  \max_{\tau_j \in \Gamma_{\xi in}^{o} } {L_j} - 1
\end{equation}

Eq. \ref{eq:maxbuffer} states that maximum occupation of the packet buffer in switch $\xi$ is the size of the payload ($L_j - 1$) of the largest packet injected into that switch, as incoming flits are only buffered when an injection is ongoing and therefore the header of the injected packet must have already exited the switch one cycle earlier during an idle tick.  

This leads to a tighter formulation for $I_i^{pos}$ by adding up the $B^o_\xi$ of every switch along the downstream path of $\tau_i$:

\begin{equation}\label{eq:ipos_1}
I_i^{pos} \ =  \sum_{\xi \in dpath^o(\pi_i^s,\pi_i^d)} {B^o_\xi}
\end{equation}

We can now calculate the total worst-case latency experienced by any packet of $\tau_i$ by adding its no-load latency $C_i$ to the worst-case interference it can suffer before and after injection:

\begin{equation}\label{eq:ribasic}
R_i \ =  C_i +  I_i^{pre} + I_i^{pos}
\end{equation}

\subsection{Indirect Interference Jitter}\label{indirect}

In this subsection, we address the issue of assigning a safe value to the indirect interference jitter $J_j^k$ of each direct interfering flow $\tau_j \in \Gamma_{up_i}^{o}$ in the third term of Eq. \ref{eq:ipre}. As explained earlier, $J_j^k$ represents the increased jitter that any $\tau_j$ may suffer due to upstream or injection interference by a flow $\tau_k$ that does not directly interfere with the flow $\tau_i$ under analysis. In ~\cite{Shi08}, Shi and Burns observed that the amount of indirect interference jitter suffered by a flow $\tau_j$ was bounded by $R_j - C_j$: the jitter can never be larger than the difference between $\tau_j$'s worst-case latency and its maximum no-load latency. Whilst potentially pessimist, that bound is safe and has been used in subsequent analyses such as ~\cite{Nikolic2018} and ~\cite{Xiong17}. However, in that approach the worst-case latency $R_j$ of a flow must be calculated before the calculation of the worst-case latency $R_i$ of any flow $\tau_i$ that suffers direct interference from $\tau_j$. In a priority-based system such as the one considered in ~\cite{Shi08}, it is always possible to calculate the worst-case latency $R_j$ of higher priority flows before using those values in the calculation of the worst-case latency $R_i$ of lower priority flows. However, in the case of a routerless NoC where flows can both interfere and suffer interference from one another in different stages of their lifetimes, such an approach becomes infeasible.

A simple solution to this problem is to assume $D_j - C_j$ as the upper bound for the indirect interference jitter $J_j^k$ of a given flow, assuming that it is schedulable. In that case, $R_j$ will not be larger than $D_j$, so the bound is safe. However, it is likely to be a very pessimistic bound. Specially in cases of underloaded networks, where packet flows are more likely to have a worst-case latency which is much smaller than their respective deadlines.

\begin{algorithm}[H]
\caption{Iterative Schedulability Test}
\label{alg:IUP}
\KwResult{Worst-case latency $R_i$ for every $\tau_i \in \Gamma$ if flowset $\Gamma$ is schedulable, $\emptyset$ otherwise}

  \SetKwInOut{Input}{inputs}
  \SetKwInOut{Output}{output}
  \SetKwProg{Schedulability}{Schedulability}{}{\textbf{end}}
  \SetKwRepeat{Do}{do}{while}

 \Schedulability{$(\Gamma)$}{
    \Input{set $\Gamma$ of traffic flows, pre-calculated values of $I^{pos}_i$ for each flow }
    \Output{set $R := \{ R_i \: | \: \tau_i \: \in \: \Gamma \}$ of worst-case latencies $R_i$ for each flow}

    \ForEach{$(\tau_i \in \Gamma)$}{
        $R_i \gets 0$\;
        $J^k_i \gets 0$\;
        
        }
    \Do{$change==true$}{
        $change \gets false $\;
        \ForEach{$(\tau_i \in \Gamma)$}{
        calculate $I_i^{pre}$ by feeding the latest values of $J^k$ into Eq. \ref{eq:ipre}\;
        calculate $R_i^{new} = C_i + I_i^{pre} + I_i^{pos}$\;
         \If{$R_i^{new} > D_i$}{ $\Gamma$ is unschedulable, return $R \gets \emptyset$ and exit\; }
         \If{$R_i^{new} \ne R_i$}{ 
            $change = true$\;
            $R_i \gets R_i^{new}$\;
            $J^k_i = R_i - C_i$\;
         }
        }
    
    }
    
     \KwRet{$R$}\;
 }
\end{algorithm}

\vspace{12pt}

In this paper, we propose a slightly more sophisticated approach to calculate $J_j^k$. We initially assume that $J_j^k = 0$, which allows for the calculation of initial (and potentially unsafe) values of $R_i$ of all flows in $\tau_i \in \Gamma$ using only the parameters available in the system model described in subsection \ref{system_model}. Once we obtain values of $R$ for individual flows, we use those to update the corresponding values of $J_j^k$ with $R_j - C_j$ (as in ~\cite{Shi08}) and use them on the calculations of the worst-case latency $R$ of other flows. As we keep iterating over those steps, the recalculated values of $R$ can increase or remain the same within an iteration, but will never decrease. So we simply keep repeating the monotonic process until the recalculated values of $R$ of all flows remain the same over two subsequent iterations (meaning that the flowset is schedulable), or at least one of them exceeds the deadline of the respective flow (meaning that the flowset is unschedulable). We present the pseudo-code of such approach as Algorithm \ref{alg:IUP}.

While both approaches are safe, i.e. both provide an upper-bound latency for each and every packet of a traffic flow, the goal of our approach is to provide tighter bounds. That means employing a worst-case latency estimate that is less pessimistic, and therefore lower. In Section \ref{experiments}, we evaluate and quantify the advantages of our approach.  

This completes the basic analysis for a routerless NoC. We now extend this analysis to cover platforms that share injection and ejection links.

\subsection{Analysis of Shared Injection Link}\label{analysis_shared_injection}

Let us now remove the assumption that each ring has its independent injection link, and update our analysis so that it takes into account the potential timing interference that arises when all rings share the same injection link (which is the implementation choice in ~\cite{liu_imr_2016} and ~\cite{alazemi_routerless_2018}). Only the formulation for $I^{pre}$ must be updated, since the additional interference will only affect $\tau_i$ before its injection. 

As we argued in the previous subsection, at most one packet of each flow using the same injection link can be queued when a packet of $\tau_i$ joins the queue. However, packets entering a different ring as $\tau_i$ will suffer interference from different sources as the ones that share the same ring, and therefore will block $\tau_i$ in different ways, so our analysis must take that into account. 

To handle this problem, we first separate the two parts of $I^{pre}$, namely the time a packet spends in the injection queue (which we refer to as $I^{pre_{queue}}$) and the time it spends at the head of the queue waiting for an idle cycle on its respective ring (which we refer to as $I^{pre_{idle}}$). Following the same intuition used in Eq. \ref{eq:ipre}, we can state that $I^{pre_{idle}}$ of a given packet of flow $\tau_i$ at the head of the injection queue can be formulated as the first idle cycle in its respective ring:

\begin{equation}\label{eq:ipreidle}
I_i^{pre_{idle}} \ =  1 + \sum_{\tau_j \in \Gamma_{up_i}^{o} } { \left\lceil
{\frac{I_i^{pre_{idle}} + J_j  + J_j^k }{T_j} } \right\rceil \cdot L_j }
\end{equation}

The only difference between Eqs. \ref{eq:ipre} and \ref{eq:ipreidle} is the additional waiting in the queue represented by the second term in Eq. \ref{eq:ipre}, which is absent in Eq. \ref{eq:ipreidle}.

We now can define $I^{pre_{queue}}$ of a packet of flow $\tau_i$ as the sum of all the $I^{pre_{idle}}$ of the packets $\tau_j$ that can be ahead of it in the injection queue:

\begin{equation}\label{eq:iprequeue}
I_i^{pre_{queue}} \ =  \sum_{\tau_j \in \Gamma_{in_i} } {(L_j + I_j^{pre_{idle}})}
\end{equation}

where $\Gamma_{in_i}$ is the subset of $\Gamma$ (and not of $\Gamma^{o}$) with all flows originating from core $\pi_i$ (i.e. the same core as $\tau_i$), regardless of which ring they use. The intuition here is that a packet of $\tau_i$ will be queued, in the worst case, for the maximum time it takes for a single packet of each of the flows sharing its injection link to enter their respective rings.

And of course the total worst-case interference before injection $I^{pre}$ under a shared injection link is given by:

\begin{equation}\label{eq:ipreshared}
I_i^{pre} \ =  I_i^{pre_{idle}} + I_i^{pre_{queue}} 
\end{equation}

Note that this formulation is general enough to be applied in the case where injection queues are not shared, i.e. when all queued packets will be injected to the same ring, so it can actually be applied instead of Eq. \ref{eq:ipre}. However, it is a more pessimistic formulation, as it always assumes the maximum time each individual packet has to wait in the head of the queue to enter the ring, while Eq. \ref{eq:ipre} considers the longest time interval with enough idle cycles for all packets in the queue to enter the ring (which can never be larger).  

Due to the separation of $I^{pre}$ into two parts, the iterative approach for the calculation of $J^k$ proposed in subsection \ref{indirect} has to be modified accordingly. Algorithm \ref{alg:IUP2} shows the changed parts from
Algorithm \ref{alg:IUP}, which mainly consists of an additional loop performed at each iteration to calculate $I^{pre_{idle}}$ for all flows before the loop used for the calculation of $I^{pre_{queue}}$, $I^{pre}$ and $R^{new}$. This is because the calculation of $I_i^{pre_{queue}}$ of a given flow depends on the values of $I^{pre_{idle}}$ of all other flows sharing the same injection link, not only its own. The remainder of the algorithm follows exactly the same logic explained in subsection \ref{indirect}.

\begin{algorithm}[tbh]
\caption{Iterative Schedulability Test for Shared Injection Links.}
\label{alg:IUP2}

        $change \gets false $\;
        \ForEach{$(\tau_i \in \Gamma)$}{
        calculate $I_i^{pre_{idle}}$ by feeding the latest values of $J^k$ into Eq. \ref{eq:ipreidle}\;
        }
        
        \ForEach{$(\tau_i \in \Gamma)$}{
        calculate $I_i^{pre_{queue}}$ by feeding the latest values of $I^{pre_{idle}}$ into Eq. \ref{eq:iprequeue}\;
        calculate $I_i^{pre} \ =  I_i^{pre_{idle}} + I_i^{pre_{queue}} $\;
        calculate $R_i^{new} = C_i + I_i^{pre} + I_i^{pos}$\;
         \If{$R_i^{new} > D_i$}{ $\Gamma$ is unschedulable, return $R \gets \emptyset$ and exit\; }
         \If{$R_i^{new} \ne R_i$}{ 
            $change = true$\;
            $R_i \gets R_i^{new}$\;
            $J^k_i = R_i - C_i$\;
         }
        }
  \end{algorithm}

\subsection{Analysis of Shared Ejection Link}\label{analysis_shared_ejection}

To reduce even further the energy and area overheads of routerless NoCs, ~\cite{liu_imr_2016} and ~\cite{alazemi_routerless_2018} consider switches with shared ejection links. In the case of Liu et al. ~\cite{liu_imr_2016}, switches only have a single ejection link shared by all rings that deliver packets to the core connected to that switch. Aware of the negative impact of a single ejection link on the NoC performance, Alazemi et al. ~\cite{alazemi_routerless_2018} supported multiple ejection links per switch, but allowed them to be shared in case the number of rings delivering packets to a given core exceeds the number of available ejection links. In both cases, when a packet arrives to its destination switch and is unable to obtain arbitration to the ejection link (i.e. a packet ejection from another ring sharing the same ejection link is already under way), it is deflected to the switch's output port for that ring and forced to do a loop around the whole ring before arriving back at the destination switch's input port and trying again to obtain arbitration to its ejection link. To prevent successive deflections from causing a livelock, both approaches propose mechanisms to limit the number of deflections: Liu et al. ~\cite{liu_imr_2016} timestamp every packet and use an Oldest-First arbitration of the ejection link; while Alazemi et al. ~\cite{alazemi_routerless_2018} tag packets with a circling counter that is incremented after every deflection, and implement a reservation scheme that guarantees an ejection link for the packet after one final loop once the counter reaches a certain value.    

Such packet deflection mechanisms are not amenable to time predictability, since they force the analysis to assume that whenever a packet can be deflected, regardless of how unlikely, it will be. While the livelock prevention mechanisms proposed by ~\cite{liu_imr_2016} and ~\cite{alazemi_routerless_2018} ensure that the number of deflections is bounded, that upper-bound may be much higher than the typical number of deflections a packet would experience (e.g. an 8-bit circling counter, as implemented in ~\cite{alazemi_routerless_2018} would allow a packet to circle the ring up to 256 times before it is guaranteed access to an ejection link). Nonetheless, we update our analysis so that it can take into account the increased worst-case packet latencies due to deflection. 

A packet deflection affects both $I_i^{pre}$ and $I_i^{pos}$ terms of the proposed analysis. The value of $I_i^{pos}$ will increase because of the additional links a packet must go through while circling around the loop after a deflection, as well as the downstream direct interference which in this case should consider traffic over the whole ring. In turn, the value of $I_i^{pre}$ will increase because of the additional packets circling the ring that can delay even further a packet injection.

We start by formulating the worst case latency a packet of a flow $\tau_i$ will experience after a deflection. We can produce two equations with different levels of precision, following the same intuition behind Eqs. \ref{eq:ipos} and \ref{eq:ipos_1}, but considering that the packet must go through all $r^o$ switches of the ring $\Xi^o$ instead of only its downstream path:

\begin{equation}\label{eq:idefl}
I_i^{defl} \ =  r^o  \cdot B^o 
\end{equation}

\begin{equation}\label{eq:idefl_1}
I_i^{defl} \ =  \sum_{\xi \in \Xi^o} { B^o_\xi}
\end{equation}

Eq. \ref{eq:iposshared} below provides an upper bound for $I_i^{pos}$ considering that a packet may suffer up to $maxloop$ deflections:

\begin{equation}\label{eq:iposshared}
I_i^{pos_{defl}} \ =   I_i^{pos} + maxloop_i \cdot I_i^{defl}
\end{equation}

It accounts for the interference $\tau_i$ may suffer from packets that can either be injected or buffered in each of the switches along its path to its destination, as well as those over each of its loops around the ring after each deflection. It provides a safe upper bound regardless if the value of $I_i^{pos}$ is obtained from Eq.~\ref{eq:ipos} or \ref{eq:ipos_1}, and the value of $I_i^{defl}$ is obtained from Eq.~\ref{eq:idefl} or \ref{eq:idefl_1}, but the tightest bound is obtained using values from Eqs.~\ref{eq:ipos_1} and \ref{eq:idefl_1}, respectively.

We now focus our attention on updating the upper bound to $I_i^{pre}$ due to the additional injection delay imposed by deflected packets. For the sake of completeness, we will present one formulation assuming an exclusive injection link per ring and another for the case where a single injection link is shared among all rings.  

If we assume that each ring has its exclusive injection link, all we need to do is to update Eq. \ref{eq:ipre} so that it also models the additional interference caused by the deflections of packets using that ring. We do that by adding a fourth term to Eq. \ref{eq:ipre}, which models the injection delay caused by each of the packet deflections by accounting for them as if they are replicas of the original packet flow, producing interfering packets with the same period, jitter and upstream indirect interference:

\begin{equation}\label{eq:ipre_2_1}
\begin{split}
I_i^{pre} \ =  1 + \sum_{\tau_j \in \Gamma_{in_i} } {L_j} + \sum_{\tau_j \in \Gamma_{up_i}^{o} } { \left\lceil
{\frac{I_i^{pre} + J_j  + J_j^k }{T_j} } \right\rceil \cdot L_j } \\
+ \sum_{\tau_j \in \Gamma^o }   \sum\limits_{1}^{maxloop_j} { \left\lceil
{\frac{I_i^{pre} + J_j  + J_j^k }{T_j} } \right\rceil \cdot L_j }
\end{split}
\end{equation}

To ensure that all possible packet deflections are considered (and not only those by the packets that share links with $\tau_i$), the outer sum in the fourth term is performed over the set $\Gamma^o$ that includes all flows using the ring $o$.

Now, let us assume that a single injection link is shared among all rings. As in subsection \ref{analysis_shared_injection}, for this case we analyse separately the time a packet spends in the injection queue ($I^{pre_{queue}}$) and the time it spends at the head of the queue waiting for an idle cycle on its respective ring ($I^{pre_{idle}}$). Deflected packets will potentially increase the worst-case wait for an idle cycle on a given ring, so we must account for that on an updated definition of $I^{pre_{idle}}$:

\begin{equation}\label{eq:ipre_2_2}
\begin{split}
I_i^{pre_{idle}} \ =  1 + \sum_{\tau_j \in \Gamma_{up_i}^{o} } { \left\lceil
{\frac{I_i^{pre_{idle}} + J_j  + J_j^k }{T_j} } \right\rceil \cdot L_j } + \\
\sum_{\tau_j \in \Gamma^o }   \sum\limits_{1}^{maxloop_j} { \left\lceil
{\frac{I_i^{pre_{idle}} + J_j  + J_j^k }{T_j} } \right\rceil \cdot L_j }
\end{split}
\end{equation}

The update follows exactly the same intuition behind Eq. \ref{eq:ipre_2_1}, and adds a term to account for packet deflections as if they are replicas of the original packet flow. The calculation of $I^{pre}$ then uses an unchanged Eq. \ref{eq:ipreshared}, adding $I^{pre_{queue}}$ provided by an unchanged Eq. \ref{eq:iprequeue} to $I^{pre_{idle}}$ obtained from Eq. \ref{eq:ipre_2_2} above.    

Finally, Eq. \ref{eq:ribasic} must be updated to include the packet's no-load latency over each of its deflections, which is accounted for by the product of its basic loop latency $C^o_i$ and the maximum number of deflections $maxloop_i$. That value is added to the packet's no-load latency from its source to destination (i.e. before any deflections take place, as in Eq. \ref{eq:ribasic}), and to the values for the worst-case interference before and after injection using the equations derived in this subsection:

\begin{equation}\label{eq:ridef}
R_i \ =  C_i + C^o_i \cdot maxloop_i  + I_i^{pre} + I_i^{pos}
\end{equation}

\subsection{Bounding Deflections}

Eqs. \ref{eq:iposshared}, \ref{eq:ipre_2_1}, \ref{eq:ipre_2_2} and \ref{eq:ridef} all require an appropriate value for the $maxloop$ variable. That value depends on the the livelock prevention mechanism used by the network, but also on the traffic load injected into it. 

In the case of the Oldest-First livelock prevention mechanism proposed by Liu et al. ~\cite{liu_imr_2016}, $maxloop$ should be set to the highest possible number of packets that can be en route to the destination core $\pi_i^d$ of the packet flow under analysis $\tau_i$, as each of them could, in the worst case, force a deflection to a packet of $\tau_i$. Since we assume that a packet flow deadline is no greater than its period, we can assume that in a schedulable system each flow will have at most one packet en route to their destination core at any given time, so the value of $maxloop$ in such a case is exactly the number of flows sharing the same destination core. 

In the routerless architecture proposed by Alazemi et al. ~\cite{alazemi_routerless_2018}, livelock is prevented by a circling counter which locks the ejection link of a ring for every packet that has reached the number of deflections given by the counter's resolution. The value of $maxloop$ could then be directly provided by the counter's resolution (i.e. its maximum value), but in reality the actual traffic load may dictate that the maximum number of deflections could be smaller (for lightly loaded networks) or even larger than the counter's resolution (as an overloaded network may force the deflection counter of specific packets to overflow, a possibility that is not addressed in~\cite{alazemi_routerless_2018}). For example, such a scenario may arise when a particular packet's circling counter expires while the ejection link is still locked to another packet. This shows that the usage of Alazemi et al.'s architecture to support real-time systems would require careful investigation of the assignment of specific values for the circling counter of each packet flow, aiming to verify feasibility and maximise schedulability resulting from that assignment. Such investigation is outside the scope of this paper and is left as future work. 

While ~\cite{liu_imr_2016} focused mostly on single ejection links shared by all flows with the same destination core, ~\cite{alazemi_routerless_2018} already supports multiple ejection links per core. By allowing for multiple ejection links while enforcing that packets of a given flow would always use the same ejection link, and by utilising an Oldest-First mechanism such as proposed in ~\cite{liu_imr_2016}, the analysis proposed in this paper enables system designers to fully exploit a well-defined trade-off between hardware resources and real-time schedulability: by adding more ejection links per core, the value of $maxloop$ can be driven down (as in this case $maxloop$ would be equal to the number of flows sharing an ejection link), and therefore schedulability becomes easier to achieve since most of the components of the flow's worst-case latency are functions of $maxloop$. A systematic exploration of this trade-off within an optimisation framework is also left as future work.

\section{Evaluation}\label{experiments}

The superiority of routerless NoCs over router-based wormhole mesh networks on average-case performance, chip area and energy dissipation has been discussed in detail in ~\cite{alazemi_routerless_2018} and ~\cite{liu_imr_2016}, supported by extensive experiments based on simulation models and hardware designs. Our goal in this section is to add a new aspect to the comparison between the two types of NoC architectures, focusing exclusively on the ability of routerless NoCs to guarantee the delivery of application communication packets within their deadlines even in the worst-case scenario. Such a comparison, however, is not necessarily straightforward. As discussed in Section \ref{analysis}, different assumptions with regards to the sharing of injection and ejection links, and therefore deflection, result in different analytical models and, most likely, different latency upper bounds. Another difficulty lies on achieving a fair comparison between routerless NoCs and router-based baselines: since their topologies and communication mechanisms are so distinct, it is likely that schedulability metrics could be impacted by the nature of the application benchmark (or its mapping onto the NoC platform) in very different ways for the routerless NoC or the baseline. For example, the same application communication traffic could flow between neighbouring cores in one of the architectures, while in the other it would require a large number of hops and potentially suffer more interference.

Taking into account the above-mentioned considerations, we performed two types of experimental evaluation. Firstly, we evaluate the ability of both types of NoC architectures to provide latency guarantees to all flows of a given flowset, i.e. full schedulability. We refer to that evaluation as \emph{flowset-based evaluation}, which is presented in detail in subsection \ref{experiments_flowset}. We then change the focus of our experimental evaluation to individual flows within a flowset, aiming to better understand how different configurations of routerless NoCs affect their individual upper-bound latencies. We refer to that evaluation as \emph{flow-based evaluation}, with the results presented in subsection \ref{experiments_flow}.

\subsection{Flowset-based Evaluation}\label{experiments_flowset}

To account for the significant difference between mesh-based and routerless NoCs, we focus first on a style of evaluation that relies on a diverse set of synthetically generated benchmarks that are not biased to any of the architectures we are comparing. Let us describe such evaluation style in more detail by stating its experimental metrics and goals:

\begin{itemize}
    
    \item The main experimental metric considered in this subsection is the \emph{schedulability ratio}, which we define as the percentage of cases out of a set of benchmarks that are deemed fully schedulable by a particular approach. If the set of benchmarks is sufficiently large and diverse, we can argue that in a comparison among multiple approaches it is the one with the highest schedulability ratio that would be more likely to produce a fully schedulable outcome in a practical NoC deployment. 
    
    \item We aim to evaluate the change in schedulability ratio over different levels of application communication load, aiming to provide more diversity to our set of benchmarks and to obtain a better comparison between different approaches under challenging operating conditions.
    
    \item We aim to investigate the impact of sharing injection and ejection links in the schedulability ratio of routerless NoCs. 
    
    \item We aim to investigate the impact of different upper bounds for route deflection in the schedulability ratio of routerless NoCs with shared ejection links.
    
    \item We aim to investigate the impact of different maximum packet sizes in the schedulability ratio of routerless NoCs.

\end{itemize}

In the remainder of this subsection, we will use the schedulability ratio metric to compare different configurations of a routerless NoCs against a state-of-the-art router-based wormhole NoC, which we will refer as the experimental \emph{baseline}. The chosen baseline follows the widely used QNoC template proposed in ~\cite{Bolotin04}: its cores are organised over a bi-dimensional grid, each of them connected to a router supporting virtual channels with flit-level priority-preemptive arbitration, and each router connected to its neighbours via bi-directional links in a 2D mesh topology. Its configuration includes XY dimension-order routing, credit-based flow control (to allow flits to cross a link in a single clock cycle), and 2-position FIFO buffers per virtual channel (as recommended in ~\cite{Indrusiak2018}, to minimise the effects of MPB). To calculate the schedulability ratio for the baseline, we will rely on the state-of-the-art analysis~\cite{Nikolic2018}.

The routerless NoC chosen for the comparison will have the same number of processing cores placed over a similar bi-dimensional grid, they will be connected by a multi-ring topology produced by Alazemi et al.'s RLrec algorithm ~\cite{alazemi_routerless_2018} (as shown in Figure \ref{fig:overview} for a 4x4 NoC, in which case the algorithm produces 10 rings). Such topology allows packets to reach a given destination through multiple rings, so we will consider a network with interfaces that choose the ring that can reach the packet's destination core with the lowest number of hops. To cover the whole scope of our contribution in this paper, we will consider several configurations of routerless NoCs regarding the sharing of injection and ejection links. 

To compare all the routerless NoC configurations against each other and the baseline, we produced a wide variety of benchmarks, each of them consisting of 100 randomly generated flowsets. To cover a wide range of communication load levels, benchmarks are characterised by the number of sporadic traffic flows composing each of their flowsets, and by the packet sizes that those flows are allowed to transfer over the network. All experiments start with a benchmark containing 100 flowsets of 20 traffic flows each, and we compare the schedulability ratio of all proposed approaches against each other and the baseline for that benchmark, i.e. how many of the 100 flowsets can be made fully schedulable by using each of the configurations (i.e. every single traffic flow in a flowset is schedulable). We then continue to generate 100-flowset benchmarks, but with more flows per flowset, and perform the same comparison for each benchmark. We stop either at benchmarks with 400 flows per flowset, or at the stage that no further comparative analysis is possible because only a single configuration is still able to provide fully schedulable flowsets (i.e. non-zero schedulability ratio). 

To produce realistic benchmarks, all sporadic traffic flows have parameters uniformly sampled from the same ranges: periods between 1 and 100 microseconds, and release jitters between 0 and 50\% of the respective periods. In each experiment, we randomly map all traffic flows over a NoC architecture in such a way that the source and destination cores are distinct, so that all flows use the NoC interconnect. We perform experiments using two different NoC sizes (4x4 and 5x5 cores, both set to operate at a clock frequency of 1 GHz) and for four different ranges of packet sizes ($L_i$ uniformly distributed between 16-48, 32-96, 48-256 and 96-512).  

\subsubsection{Independent ejection links}\label{experiments_flowset_IE}

The first flowset-based evaluation we present here assumes that all routerless NoC configurations have independent ejection links, and therefore there is no need for packet deflection (i.e. flows will always have an ejection link available to them). This requires larger hardware overheads, but poses a smaller impact on performance and performance predictability. The aims of this comparison are (i) to compare such routerless NoC configurations against a mesh-based priority-preemptive baseline, (ii) to investigate the impact of sharing or not the injection links to the NoC, and (iii) to evaluate the advantages of using the proposed iterative algorithms to calculate the indirect interference $J^k$ (described in Algorithms \ref{alg:IUP} and \ref{alg:IUP2}) against the simplified calculation of $J_j^k = D_j - C_j$.

\begin{sidewaysfigure}
 \centering
  \includegraphics*[scale=0.475]{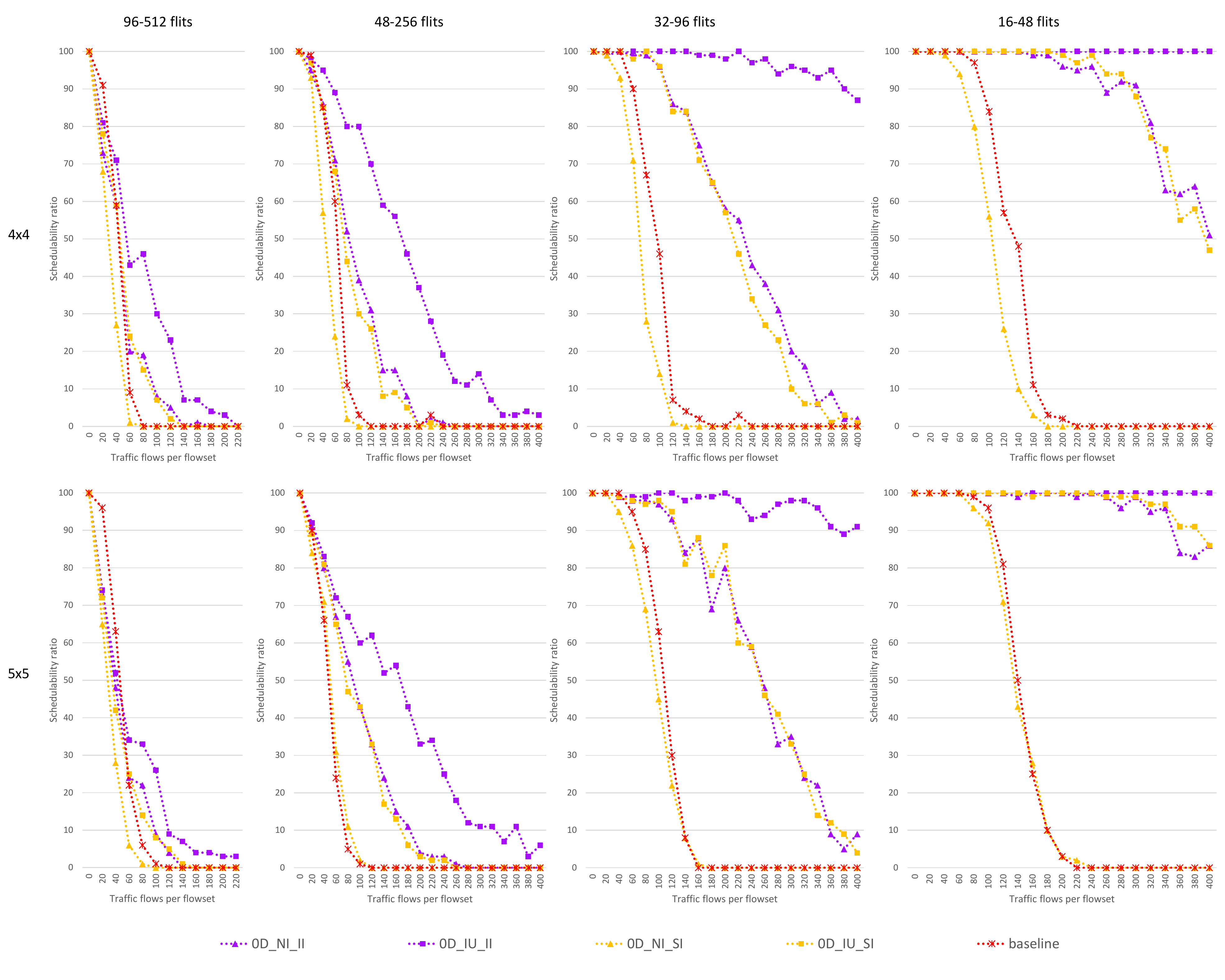}
  \caption{Flowset-based evaluation of routerless NoCs with independent ejection links against a priority-preemptive mesh baseline}
  \label{fig:expIE}
\end{sidewaysfigure}

Figure \ref{fig:expIE} shows eight plots, covering the two NoC sizes and four ranges of packet sizes. Within each plot, there are five curves displaying the schedulability ratio achieved for each benchmark by the following NoC configurations:

\begin{itemize}
\item 0D\_NI\_II - routerless NoC with independent ejection, non-iterative calculation of indirect interference jitter (i.e. $J_j^k = D_j - C_j$), independent injection.
\item 0D\_IU\_II - routerless NoC with independent ejection, iterative calculation of indirect interference jitter (i.e. Algorithm \ref{alg:IUP}), independent injection.
\item 0D\_NI\_SI - routerless NoC with independent ejection, non-iterative calculation of indirect interference jitter (i.e. $J_j^k = D_j - C_j$), shared injection.
\item 0D\_IU\_SI - routerless NoC with independent ejection, iterative calculation of indirect interference jitter (i.e. Algorithm \ref{alg:IUP}), shared injection.
\item baseline - mesh-based NoC with priority-preemptive arbitration 
\end{itemize}

A careful analysis of the plots in Fig. \ref{fig:expIE} allowed us to confirm the following experimental hypotheses:

\begin{itemize}

\item \emph{The proposed iterative algorithm for the calculation of indirect interference jitter provides tighter upper-bounds than the simplified non-iterative method}. This can be seen in both NoC sizes, and across all ranges of packet sizes. For the case of independent injection, the schedulability ratio of the configuration 0D\_NI\_II (analysed using the simplified bound) is consistently worse than the 0D\_IU\_II (analysed using the proposed iterative algorithm). Similarly for the case with shared injection, 0D\_NI\_SI is consistently worse than 0D\_IU\_SI. This can be clearly seen in the plots: when comparing lines of the same colour, the ones with triangle markers (simplified) are never above the ones with square markers (iterative). In the cases with smaller packets, where the network is less saturated and therefore worst-case latencies are more likely to be well below the respective deadlines, the advantages of the proposed iterative analysis are much more significant (as the simplified bound becomes very pessimistic, as explained in subsection \ref{indirect}). For example, in the 4x4 NoC with packets between 16 and 48 flits and shared injection links (i.e. yellow lines), the iterative approach (square marker) produced a 100\% schedulability ratio for benchmarks with up to 180 flows per flowset, while the simplified method (triange marker) dropped below 100\% schedulability ratio already for benchmarks with 40 flows per flowset, and at 180 flows per flowset its schedulability ratio dropped to zero. Given the consistent superiority of the iterative analysis, which can be observed with NoC sizes, packet sizes and injection link sharing status, we will focus the subsequent comparisons on the configurations analysed using the iterative approach (i.e. 0D\_IU\_II and 0D\_IU\_SI).

\item \emph{Sharing injection links has detrimental impact on worst-case performance}. This was of course expected, as sharing injection with packets going into different rings may introduce much larger worst-case injection latencies. The evaluation results in Fig. \ref{fig:expIE} allow us to quantify that impact by measuring the vertical distance between 0D\_IU\_SI (shared injection) and 0D\_IU\_II (independent injection) in the plots. The configuration with shared injection is consistently below the one with independent injection across all plots, typically by 5-25 percent points, but in some cases the difference can reach more than 80 percent points. For example, in the 5x5 NoC with packets between 32 and 96 flits and benchmark with 400 flows per flowset, the schedulability ratio with independent injection is 91\% while sharing the injection link allowed for a schedulability ratio of only 4\%.

\item \emph{The performance of Routerless NoCs suffers significantly as packet sizes increase}. The plots clearly show that, for the case of small packets (i.e. 16-48 and 32-96 flits), the schedulability ratio of the Routerless NoC configurations is significantly better than the baseline, for both independent and shared injection links. However, with the increase of packet sizes, that advantage shrinks and in the case with the largest packets the baseline becomes superior (particularly in the 5x5 NoC case). This is due to the fact that routerless NoCs have no means to preempt large packets, so they can impose large amounts of interference to other packets (potentially more urgent ones), simply because they were earlier in acquiring a share resource. The baseline uses preemptive arbitration, allowing packets with shorter deadlines to be given higher priority, so they can acquire a shared resource previously acquired by a large packet, preventing long undesired delays. This effect can also be seen in the plots, as the performance of the baseline does not change so widely across different packet sizes.
\end{itemize}

To summarise our experimental conclusions so far, we can firstly state that the proposed iterative approach for calculating indirect interference jitter provides significantly tighter upper bounds compared to the simplified non-iterative approach, and therefore should be used as the best known representation of the worst-case behaviour of routerless NoCs. Then, by analysing the results produced by the proposed iterative approach, we can state that a routerless NoC with independent ejection and small packet sizes is consistently superior to a mesh-based priority-preemptive NoC when it comes to real-time schedulability, even in the cases with shared injection link. For larger packets, routerless NoCs perform significantly worse and can be outperformed by mesh NoCs, so such configurations should be avoided. This can be done by exploring the trade-off between period and packet sizes, i.e. by breaking packet into smaller units that can be injected more frequently, but a detailed analysis of such optimisation approach is left as future work.   

As reviewed in subsection \ref{analysis_shared_ejection}, the cost of independent ejection links can be prohibitive and routerless NoCs typically share them among multiple rings, forcing packets to be deflected if the ejection link is in use upon their arrival. In the next subsection, our evaluation focuses in that kind of scenario.

\subsubsection{Shared ejection links}\label{experiments_flowset_SE}

Our second flowset-based evaluation aims to quantify the impact of sharing ejection links on the real-time schedulability of routerless NoCs. Of course, that impact depends on the maximum number of deflections that a packet may have to do, so our comparison will cover the following alternatives:  

\begin{itemize}
\item 0D\_IU\_SI - routerless NoC with independent ejection, iterative calculation of indirect interference jitter (i.e. Algorithm \ref{alg:IUP}), shared injection. This is the same as the configuration with the same name in the previous subsection.
\item 1D\_IU\_SI - routerless NoC with shared ejection and a maximum of 1 deflection, iterative calculation of indirect interference jitter (i.e. Algorithm \ref{alg:IUP}), shared injection. 
\item 2D\_IU\_SI - routerless NoC with shared ejection and a maximum of 2 deflections, iterative calculation of indirect interference jitter (i.e. Algorithm \ref{alg:IUP}), shared injection.
\item 3D\_IU\_SI - routerless NoC with shared ejection and a maximum of 3 deflection, iterative calculation of indirect interference jitter (i.e. Algorithm \ref{alg:IUP}), shared injection. 
\item baseline - mesh-based NoC with priority-preemptive arbitration. This is the same as the configuration with the same name in the previous subsection. 
\end{itemize}

Configuration 0D\_IU\_SI was used again in this evaluation, as it has outperformed the baseline in the previous evaluation while requiring lower hardware resources (as it uses shared injection links). We then evaluate different levels of ejection link sharing by limiting the number of deflections allowed to all packets in the NoC, from 0 to 3.

\begin{sidewaysfigure}
  \centering
  \includegraphics*[scale=0.475]{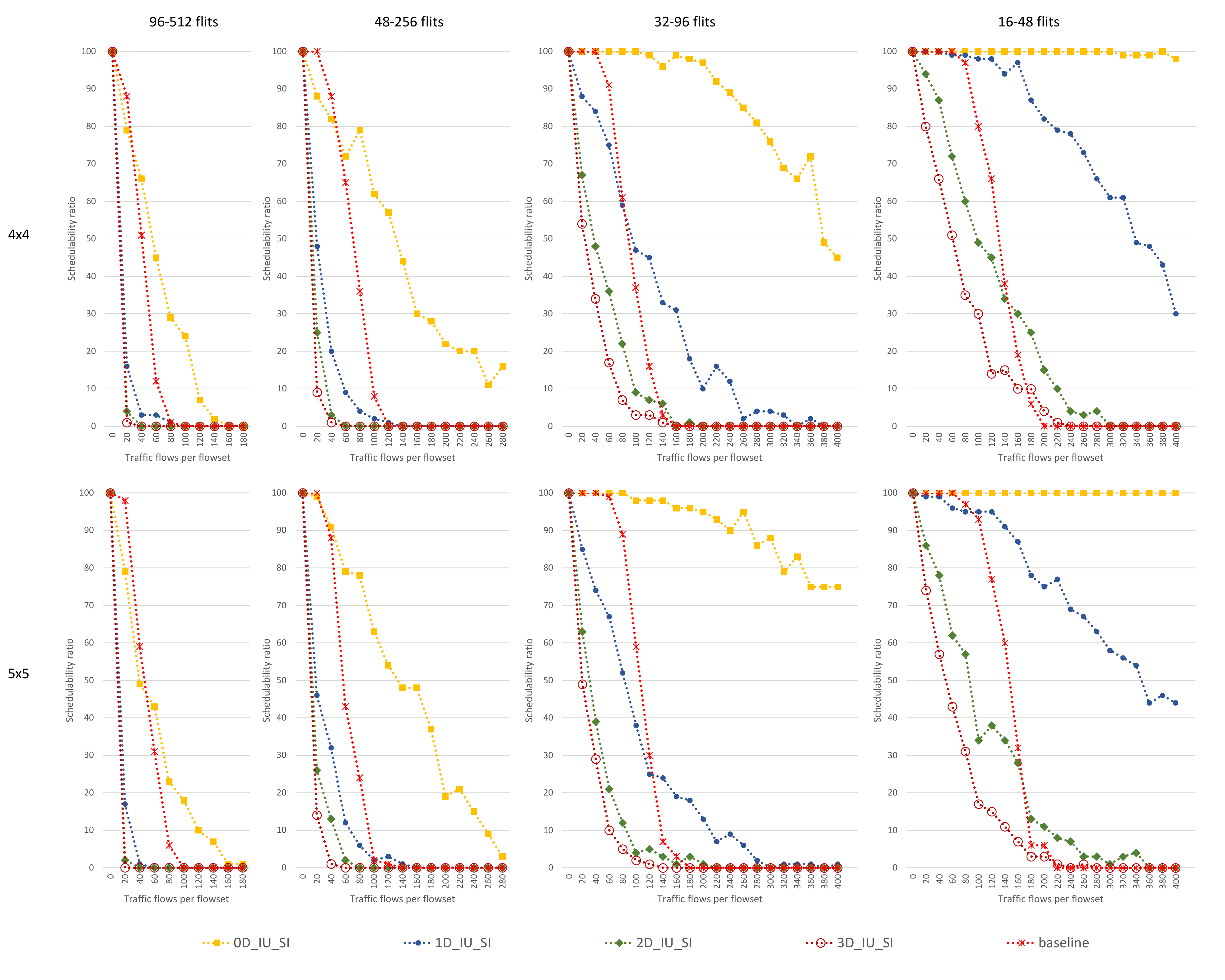}
  \caption{Flowset-based evaluation of routerless NoCs with shared ejection links against a priority-preemptive mesh baseline.}
  \label{fig:expSE}
\end{sidewaysfigure}

The generation of benchmarks follows the same process described in the previous subsection, and the NoC sizes, topologies and packet sizes are also the same.

By looking at the plots in Fig. \ref{fig:expSE}, we can see that a larger number of deflections leads to lower schedulability ratios, as expected. We can also see that the difference in schedulability ratio for each additional deflection is larger in the 5x5 NoC than in the 4x4 NoC, which can be explained by the need of deflected packets to cycle around larger rings (but which poses interesting questions regarding the scalability of routerless NoCs with shared ejection).

Once more, we can see the significant impact of packet sizes. For the smallest packets, we can see that a routerless NoC that allows up to 2 deflections for every packet can still beat the baseline (e.g. for benchmarks with more than 160 flows per flowset on a 4x4 NoC, and 180 flows per flowset on a 5x5 NoC). For larger packets, allowing even a single deflection would put the routerless NoC configurations below the baseline. 

As a summary, we can state that deflections should be seen as the exceptional case when it comes to supporting real-time guarantees in routerless NoCs. While full schedulability could be achieved when allowing up to two deflections per packet, our experimental evidence shows that those cases are not common at high levels of communication load. The real-time analysis framework proposed in this paper can enable an optimisation process that considers partitioning of flows over multiple shared ejection links, aiming to fulfil real-time guarantees while minimising the number of ejection links (and the associated hardware overhead), which we will tackle as future work.

\subsection{Flow-based Evaluation}\label{experiments_flow}
In the previous subsection, we used schedulability ratio as the metric to compare different NoC configurations and analysis techniques over a wide variety of flowsets. We now present results of a flow-based evaluation, which instead focuses on the worst-case latencies of individual flows within a flowset. Our aim is to better understand the impact of the different components of the worst-case latency under different scenarios and different levels of load. 

The experimental setup is similar to what we used in the previous subsection. We generate flowsets with different number of flows in each of them, aiming to cover a wide range of communication loads. Given the nature of the proposed analysis, we are only able to obtain valid worst-case latencies for schedulable flows (as seen in Algorithms \ref{alg:IUP} and \ref{alg:IUP2}), so the comparisons in this subsection must consider only fully schedulable flowsets. Therefore, the process of benchmark generation had to be changed: we would repeatedly apply the process described in the previous subsection for generating random flowsets within the parameters specified for the specific benchmark, stopping only when it finds a fully schedulable one, which is then used in the experimental comparisons described below.

Even though our goal in this subsection is to analyse worst-case latencies of individual flows within a flowset, it is not easy to make sense of those values directly. Instead, we focus on the spread and skewness of the latencies of flows within a flowset, and we plot that distribution as a box-and-whiskers plot: the lower and upper whiskers represent the minimum and maximum values, while the box represents the median, and the first and third quartiles.  

Following our findings from the flowset-based experiments, we now focus only on traffic flows with packets between 16 and 48 flits. In all experiments, routerless and baseline NoCs have the same size (4x4).

\subsubsection{Independent ejection links}\label{experiments_flow_IE}
As before, we start our comparative evaluation by considering the case where all rings have their own ejection link, so deflection is not necessary. Our goals here are similar to that of subsection \ref{experiments_flowset_IE}, but now looking at the impact on individual flow latencies instead of overall schedulability ratio.

\begin{figure*}[tbh]
  \centering
  \includegraphics*[scale=0.5]{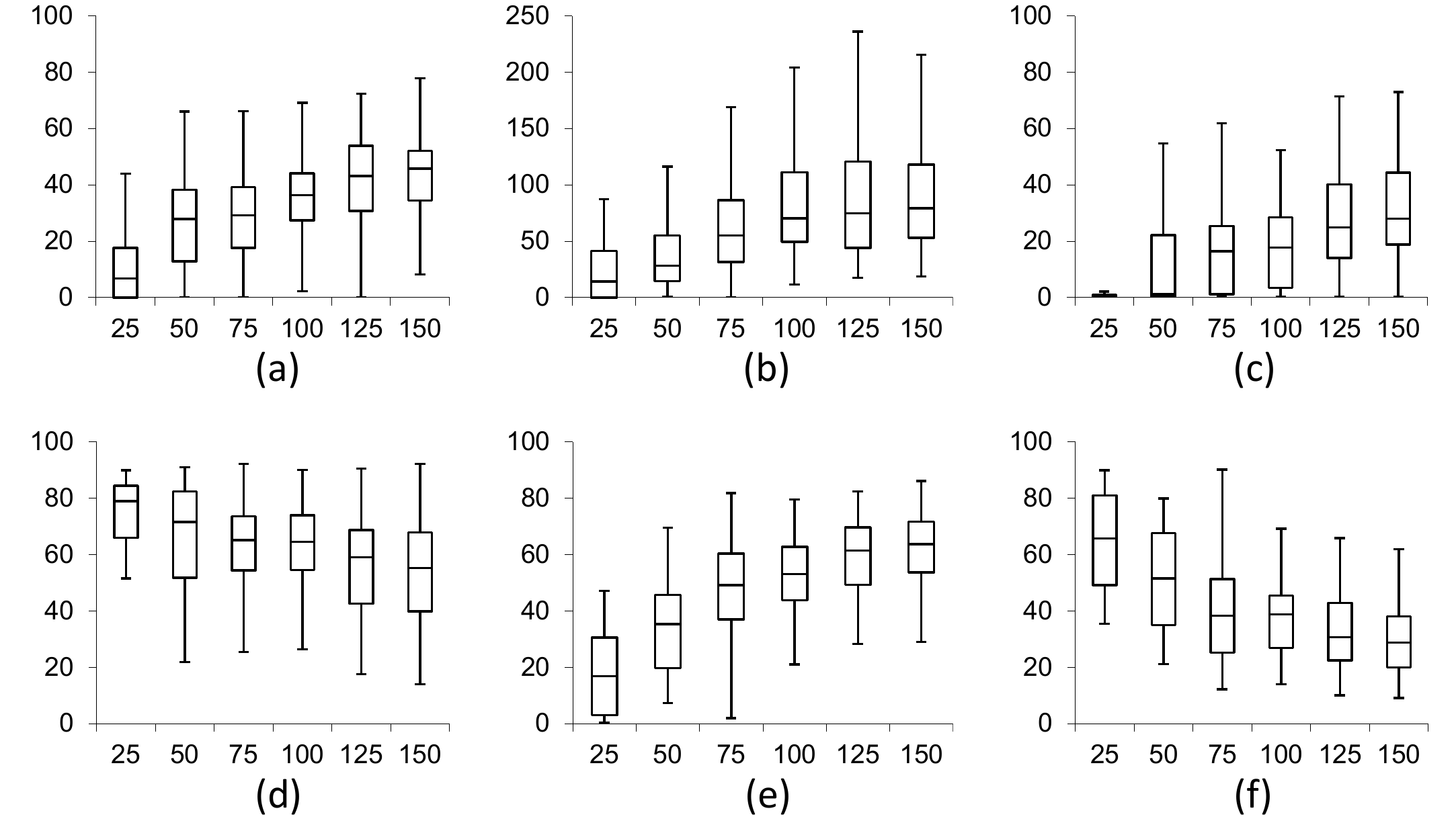}
  \caption{Flow-based evaluation of routerless NoCs with independent ejection links: x-axis shows the number of flows in the flowset, y-axis shows (a) percent difference between non-iterative and iterative solutions, (b) percent difference between shared and independent injection, (c) percentage of the worst-case latency due to $I_{pre}$ under independent injection, (d) percentage of the worst-case latency due to $I_{pos}$ under independent injection, (e) percentage of the worst-case latency due to $I_{pre}$ under shared injection, (f) percentage of the worst-case latency due to $I_{pos}$ under shared injection.}
  \label{fig:expflIE}
\end{figure*}

Let us first look into the impact of the proposed iterative algorithms to calculate the indirect interference $J^k$ against the simplified calculation of $J_j^k = D_j - C_j$. In Fig. \ref{fig:expflIE}(a), we see box-and-whiskers for six different flowsets, all fully schedulable, but each with a different number of flows (from 25 to 150, indicated on the x-axis). The data distribution represented by each box-and-whiskers is the percent difference between the worst-case latency of a given flow obtained through the proposed iterative analysis and the worst-case latency of the same flow using the non-iterative simplified calculation. For the smallest flowset with 25 flows, we see that there is no difference for up to a quarter of the flows, and for more than three quarters of the flows the proposed iterative algorithm provided worst-case bounds that were less than 20\% smaller than the simplified calculation. However, as the number of flows per flowset increases to 50, forcing more resource sharing and therefore more interference, the tighter formulation of the proposed algorithm provides a sharp improvement, with half of the flows showing an improvement above 30\%. As the number of flows per flowset increases even more, the network becomes more and more congested and therefore the advantages of the proposed approach reach a plateau: as the worst-case latency $R$ of many flows become closer to their respective deadlines, the simplified use of the deadline as a proxy for $R$ introduces less and less error. If we could keep increasing the number of flows per flowset towards network saturation, we would likely see the percent difference decrease from that plateau, but that becomes increasingly difficult because we are less and less likely to find fully-schedulable flowsets at that level of network load.

Now, let us evaluate the impact of having shared injection links. In Fig. \ref{fig:expflIE}(b), we present box-and-whiskers showing the percent difference between the worst-case latency of a flow using a routerless NoC with independent injection links for each ring, and the worst-case latency of the same flow using a routerless NoC where all rings share the same injection link on a switch. The first feature we should notice is the magnitude of the spread of the distribution, which shows that sharing the injection link can lead to an increase of the worst-case latency by more than 200\%. The increase of the median difference as we increase the network load is significant, but expected as there will be likely more congestion over the shared link. More importantly, there is a sharp increase in the maximum percent difference (i.e. upper whisker), which shows that sharing injection links is much more likely to create worst-case outliers. For a flowset to be deemed unschedulable, we only need one flow to be unschedulable, and we can now see that sharing injection links makes that more likely. This fully corroborates our findings from the flowset-based experiments, where we see a much larger gap between yellow (shared injection) and purple (independent injection) lines in Fig. \ref{fig:expIE} as the network load increases (either by increasing the number of flows or the size of packets).

In Fig. \ref{fig:expflIE}(c) we have the percentage of the worst-case latency of each flow of the flowset that is due to the worst-case latency before injection $I^{pre}$, and in Fig. \ref{fig:expflIE}(d) the percentage due to the worst-case latency after injection $I^{pos}$, both in a routerless NoC with independent injection. We can see that for small flowsets the overall latency is dominated by $I^{pos}$, as the calculation of $I^{pos}$ does not depend on the number of flows (from Eq. \ref{eq:ipos} we can see it is a function of the path length and maximum packet size). The calculation of $I^{pre}$, on the other hand, is based on sums over interference sets, which become larger as the number of flows increase. This is clearly reflected on the increase of the contribution of $I^{pre}$ and the reduction of the contribution of $I^{pos}$ as the network load increases. Nonetheless, $I^{pos}$ dominates across the the whole range of benchmarks.

In Fig. \ref{fig:expflIE}(e) and (f), we have exactly the same comparisons as in (c) and (d), but for a routerless NoC with shared injection. Sharing injection links does not impact the calculation of $I^{pos}$, because once injected a packet will still suffer interference of at most one packet per hop. However, the shared injection boosts $I^{pre}$ significantly due to the additional queuing behind packets entering other rings. In this case, we can see a breaking point for flowsets with more than 50 flows, where $I^{pre}$ becomes the dominant component of a flow's worst-case latency. This is a particularly interesting finding, as our analytical framework allows network designers to pinpoint when the injection links become the main responsible for the worst-case packet behaviour in a routerless NoC, and therefore decide at which point the price of exclusive injection links becomes worth paying.

\subsubsection{Shared ejection links}\label{experiments_flow_SE}

\begin{figure*}[tbh]
  \centering
  \includegraphics*[scale=0.5]{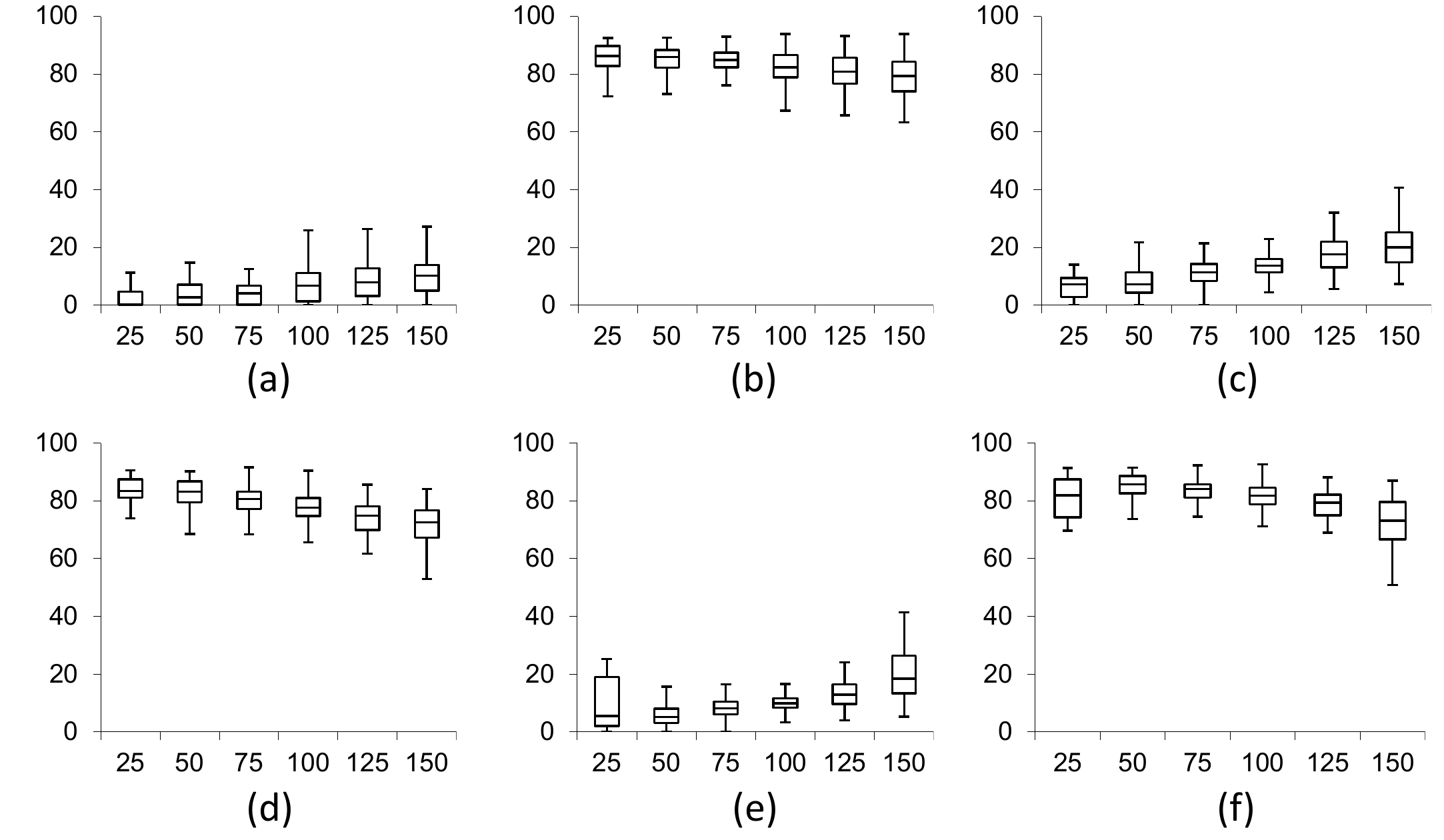}
  \caption{Flow-based evaluation of routerless NoCs with shared ejection links: x-axis shows the number of flows in the flowset, y-axis shows (a) percentage of the worst-case latency due to $I_{pre}$ with 1 deflection per packet, (b)  percentage of the worst-case latency due to $I_{pos}$ with 1 deflection per packet, (c) percentage of the worst-case latency due to $I_{pre}$ with 2 deflections per packet, (d)  percentage of the worst-case latency due to $I_{pos}$ with 2 deflections per packet, (d) percentage of the worst-case latency due to $I_{pre}$ with 3 deflections per packet, (e)  percentage of the worst-case latency due to $I_{pos}$ with 3 deflections per packet.}
  \label{fig:expflSE}
\end{figure*}

To complete our flow-based evaluation, we now look into routerless NoCs with shared ejection links and analyse the impact of deflection to the components of a flow's worst-case latency. We consider that injection links are also shared, so we can consider this evaluation as a further step beyond the comparison presented in Fig. \ref{fig:expflIE}(e) and (f).

In Fig. \ref{fig:expflSE} (a) and (b) we have again, respectively, the percentage of the worst-case latency of each flow that is due to $I^{pre}$ and $I^{pos}$, but this time with a level of ejection link sharing that allows for at most one deflection per packet (i.e. \emph{maxloop = 1}). We can see that a single deflection already reverts the trend we have seen in \ref{fig:expflIE}(e) and (f) where $I^{pre}$ became dominant for heavier network loads. If deflections are necessary, $I^{pos}$ becomes dominant across all levels of load. And since $I^{pos}$ does not depend on the number of flows in the flowset, we can see that the spread of the distribution of latencies is much smaller than what we have seem in the previous subsection.

Finally, in Fig. \ref{fig:expflSE} (c) and (d) we see the same comparison for up to two deflections per packet, and in Fig. \ref{fig:expflSE} (e) and (f) for up to three deflections per packet. Perhaps counter-intuitively, we can see that with an increased number of deflections we see a decrease in the dominance of $I^{pos}$, and an increased variability on both latency components. This is due to the fact that deflected packets have a large contribution to $I^{pre}$ (as denoted by the last term of both Equations \ref{eq:ipre_2_1} and \ref{eq:ipre_2_2}), because their reappearance will increase even more the wait for the idle network cycle that allows a new packet to be injected. Similarly, but not as significantly, the number of deflections also increases the part of the worst-case latency that is not due to $I^{pos}$ nor $I^{pre}$, i.e. the time the packet is moving uninterruptedly around the ring.

\section{Conclusions and Future Work}

This paper has identified the potential of routerless networks-on-chip as a communication backbone for real-time multiprocessor platforms, and has proposed the first real-time analysis framework for those networks. The proposed framework supports multiple configurations within the architectural space of routerless NoCs, including sharing of injection and ejection links, and the use of starvation avoidance mechanisms to bound the number of packet deflections. Evaluation using a large number of synthetic benchmarks allowed for the comparison among multiple routerless NoC configurations, and against a state-of-the-art router-based wormhole NoC baseline. We found that, besides the previously reported advantages in chip area, energy dissipation and average-case performance, routerless NoCs can be competitive alternatives when it comes to providing performance guarantees to real-time applications and systems. Our experiments have shown that this is particularly true for configurations with small maximum packet sizes and limited sharing of injection and ejection links.  

There are several limitations to the approach presented here, but also several new opportunities enabled by it, all of which provide scope for further work. Our evaluation work is based on synthetically generated benchmarks which are randomly mapped onto the NoC platforms. While this is one way to achieve fairness, i.e. the mapping is not biased towards any of the routerless configurations nor to the baseline, it fails to identify which platform is actually more amenable to optimisation. Additional experiments based on the use of metaheuristics to optimise the mapping of the application benchmarks onto each specific platform would be one way to complete the picture and evaluate the optimisation potential of the different alternatives (specially as this has already been shown to be true for the baseline, e.g. with branch-and-bound~\cite{Khan2018} and genetic algorithms~\cite{Sayuti13}). 

We have shown that the proposed analytical framework can quantify the impact of maximum packet sizes and sharing of injection and ejection links. Such framework can be used as a fitness function within the process of design space exploration for real-time embedded systems (such as in ~\cite{Penny19} and ~\cite{Ma14}). Guided by the proposed framework, a good design space exploration process would be able to evaluate alternative buffer sizes as well as the number of injection and ejection links per switch, aiming to configure a NoC platform with minimum hardware overheads while still satisfying all the real-time requirements of a given application.



\vspace{12pt}

\bibliographystyle{ieeetr}
\bibliography{refs}

\end{document}